\documentclass[journal]{IEEEtran}

\usepackage[cmex10]{amsmath}
\usepackage{cite}
\usepackage{url}
\usepackage{color}

\usepackage{amssymb}
\usepackage{path}
\usepackage{verbatim}
\usepackage{algorithm}
\usepackage{framed}
\usepackage{algpseudocode}

\usepackage{color}
\usepackage{amsmath}
\usepackage{cases}
\usepackage{theorem}
\usepackage{subfig}
\usepackage{subfloat}

\usepackage{graphicx}

\usepackage{pgfplots}
\usepackage{pgfplotstable}
\pgfplotsset{compat=newest}

\newcommand{\R}{\ensuremath{\mathbb{R}}}
\newcommand{\N}{\ensuremath{\mathbb{N}}}

\usepackage{tikz}
\usetikzlibrary{graphs,graphs.standard,quotes}
\usepackage{pgfplots}
\usetikzlibrary{positioning}
\usetikzlibrary{fit,matrix}

{ \theorembodyfont{\normalfont} 

}

\newcounter{enumctr}

\DeclareFontFamily{U}{mathx}{\hyphenchar\font45}
\DeclareFontShape{U}{mathx}{m}{n}{<-> mathx10}{}
\DeclareSymbolFont{mathx}{U}{mathx}{m}{n}
\DeclareMathAccent{\widebar}{0}{mathx}{"73}

\begin{document}

\title{MPC-CSAS: Multi-Party Computation for Real-time Privacy-preserving Speed Advisory Systems}

\author{Mingming~Liu, Long Cheng, Yingqi Gu, Ying Wang, Qingzhi Liu and Noel E. O'Connor

\thanks{Part of this work was supported by Science Foundation
Ireland under Grant No. \textit{SFI/12/RC/2289\_P2}. (\textit{Corresponding Author}: Long Cheng).
}
\thanks{M. Liu is with the School of Electronic Engineering, Dublin City University,
Ireland (e-mail: mingming.liu@dcu.ie).}

\thanks{L. Cheng is with the School of Computing, Dublin City University, Ireland (e-mail: long.cheng@dcu.ie).}

\thanks{Y. Gu is with the Insight Centre for Data Analytics, Dublin City University, Ireland (e-mail: yingqi.gu@insight-centre.org).}

\thanks{Y. Wang is with the Institute of Computing Technology, Chinese Academy of Sciences, Beijing 100190, China (e-mail: wangying2009@ict.ac.cn).}

\thanks{Q. Liu is with the Information Technology Group, Wageningen University, The Netherlands (e-mail: qingzhi.liu@wur.nl) }

\thanks{N. O'Connor is with the School of Electronic Engineering and the SFI Insight Centre for Data Analytics at Dublin City University, Ireland (e-mail: Noel.OConnor@dcu.ie)}
}

\maketitle


\begin{abstract}
As a part of Advanced Driver Assistance Systems (ADASs), Consensus-based Speed Advisory Systems (CSAS) have been proposed to recommend a common speed to a group of vehicles for specific application purposes, such as emission control and energy management. With Vehicle-to-Vehicle (V2V), Vehicle-to-Infrastructure (V2I) technologies and advanced control theories in place, state-of-the-art CSAS can be designed to get an optimal speed in a privacy-preserving and decentralized manner. However, the current method only works for specific cost functions of vehicles, and its execution usually involves many algorithm iterations leading long convergence time. Therefore, the state-of-the-art design method is not applicable to a CSAS design which requires real-time decision making. In this paper, we address the problem by introducing MPC-CSAS, a Multi-Party Computation (MPC) based design approach for privacy-preserving CSAS. Our proposed method is simple to implement and applicable to all types of cost functions of vehicles. Moreover, our simulation results show that the proposed MPC-CSAS can achieve very promising system performance in just one algorithm iteration without using extra infrastructure for a typical CSAS. 
\end{abstract}


\IEEEpeerreviewmaketitle

\section{Introduction}

With the advances in smart vehicle technologies and Intelligent Transportation Systems (ITS), Intelligent Speed Advisory (ISA) systems have become a critical part of Advanced Driver Assistance Systems (ADASs). For both manually driven and autonomous vehicles, ISA has shown to be able to significantly improve driving safety, sustainability and efficiency~\cite{hounsell2009,tradisauskas2009,darbha2018benefits}. 

\begin{figure}[h]
	\begin{center}
		{\includegraphics[width=\columnwidth]{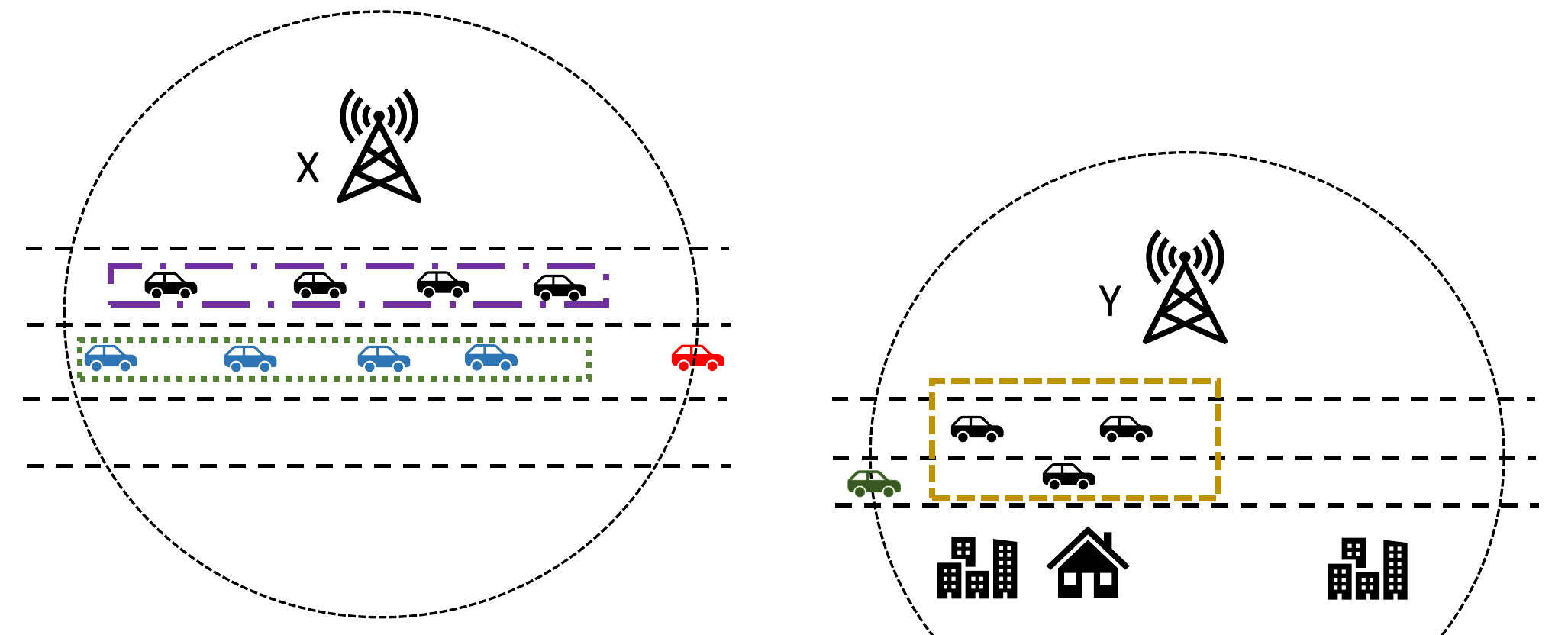}}
		\caption{An application scenarios of consensus-based speed advisory systems.}
		\label{fig:app}
	\end{center}
\end{figure}

As a special type of ISA, Consensus-based Speed Advisory Systems (CSAS) aim to recommend a consensus speed to a group of vehicles in a given area~\cite{gu2014optimised, liu2015distributed}. Compared to the cases that asking different vehicles to drive at their own optimal speeds, CSAS make more sense in many practical scenarios. As demonstrated in Fig.~\ref{fig:app}, cars tend to follow a common speed when possible in highways or special zones in cities, and this can bring some obvious benefits to various types of road users, such as reduced emissions (with less frequent accelerations/decelerations), reduced energy consumption, increased throughput, and increased safety and health~\cite{Nature,ICCVE2013,Nature2,SUMO_Optimal, gu2018design}.

Considering the condition that environmental concerns become increasingly compelling, in this paper we focus on the design of a CSAS to minimize the emissions for a group of vehicles. In fact, achieving an optimal common speed for the case is not easy. The main reason is that different vehicles are designed to operate optimally at different vehicle speeds and at different loading conditions~\cite{liu2015distributed}. Namely, to get an optimal speed, we will have to get all the detailed information of the vehicles in a group, not only the basic information such as the vehicle type, vehicle age and fuel mode, but also the load and the desired time of arrival etc. It is obvious that the data collection process will be complex. More important, due to confidentiality concerns and privacy regulations such as GDPR\footnote{\url{https://eur-lex.europa.eu/eli/reg/2016/679/oj}}, nowadays it is not always possible for vehicle owners to share or reveal their data. Therefore, a more practical and privacy-preserving CSAS becomes desirable.

Along this line, our previous work in \cite{liu2015distributed} has attempted to devise an optimal speed for a CSAS in a privacy-preserving manner, namely without revealing in-vehicle information to other vehicles or to infrastructures. Specifically, it is assumed that each vehicle is associated with an emission cost function, and only the implicit information, i.e. derivative values of the cost function at certain speeds are transmitted to a central infrastructure. For simplicity, we refer to the approach proposed in \cite{liu2015distributed} as DP-CSAS (i.e., Derivation based Privacy-preserving CSAS) in the following context.

Although DP-CSAS represents the state-of-the-art approach on privacy-preserving CSAS design, it has two shortcomings: 

\begin{itemize}
\item[(i)] The proposed optimization model in DP-CSAS is built based on the assumption that the emission cost functions of all vehicles are strictly convex, and thus the derivatives can be calculated. Further, the emission model adopted in \cite{liu2015distributed} is essentially an average-speed model, and the emission only depends on the average-speed of a specific vehicle. This unlikely works for all real cases due to (1) the emission function of a vehicle can also depend on other factors, such as acceleration, road types, weather conditions etc, and (2) some cost functions can be discontinuous and not differentiable.

\item[(ii)] DP-CSAS usually requires a relatively large amount of iterations for the algorithm convergence. This makes the optimization process time costly, and thus the applications of the approach may be limited. For example, as the results reported~\cite{liu2015distributed}, DP-CSAS takes more than 200 secs to optimize the speed for a group of 40 vehicles. It means that DP-CSAS will not be hard to handle dynamical cases in real time, such as frequent changes of a vehicular network due to frequent check-in and check-out of vehicles in the group, which may require the new optimal consensus speed to be calculated within a few seconds. 
\end{itemize}

To provide a more practical and powerful privacy-preserving CSAS, in this paper, we introduce a Multi-Party Computation (MPC) based design approach. Specifically, the proposed method, MPC-CSAS, has several advantages compared to the state-of-the-art DP-CSAS: (1) it is applicable to all emission cost functions; (2) it is simple to implement in real-time without imposing a large communication burden on the existing infrastructures; and (3) it can not only be deployed in a static and strongly connected network, but it can also be extended to deal with weakly and dynamic connectivity conditions in a practical ITS scenario under a certain assumptions. Therefore, we believe that the proposed MPC-CSAS can be applied in many new scenarios where the DP-CSAS method cannot effectively and efficiently take over.

The remainder of this paper is organized as follows. In Section~\ref{sec:relate}, we introduce the background and the related works of CSAS. We present the system design of MPC-CSAS in Section~\ref{sec:design}. We carry out extensive evaluation of our approach in Section~\ref{sec:eva} and conclude this paper in Section~\ref{sec:con}.

\section{Background and Related work}
\label{sec:relate}

In this section, we briefly introduce the optimization problem of CSAS as well as the related works.

\subsection{CSAS Problem Statement}
\label{s:Problem_Statement}

In this section, we describe an application to design a CSAS for a fleet of vehicles. Specifically, the objective of the CSAS is to recommend a consensus virtual speed to a fleet of Internal Combustion Engine Vehicles (ICEVs) running on the highway so that the overall emissions of the fleet can be minimized if all ICEVs can follow the recommended speed. Full details of this application have been presented in \cite{liu2015distributed}. Here, we briefly review the problem statement for completeness of the context. 

We consider a scenario in which a number of ICEVs are driving along a stretch of highway in the same direction. Some users of the ICEVs can decide to participate into the CSAS, that is during their driving period, they wish to use the CSAS to receive a recommend virtual speed in order to reduce the overall emissions on the highway. Each ICEV owner may get some revenue by participating into the CSAS, e.g., tax reduction. Let $N$ denote the number of ICEVs on the highway where the broadcast signal from the CSAS can be received. Each vehicle needs to equip with a specific communication device, which is able to receive the broadcast signal and transmit limited message back to the CSAS and to nearby vehicles. In practice, the broadcast signal from the CSAS can be triggered from a base station facilitated at the road infrastructure, and vehicular communication links can be established using V2V and V2I technologies.

Let $\underbar{N}:=\left\{1,2,...,N\right\}$ denote the set for indexing the vehicles, and let $s_i(k)$ be the recommend speed of the $i$'th vehicle at a time slot $k$. Let $\textbf{s}(k):=\left[s_1(k), s_2(k), \dots, s_N(k)\right]^{\textrm{T}}$ be a vector for all $s_i(k)$ at time $k$. Furthermore, each vehicle $i$ is associated with a cost function $f_i(s_i(k))$ which models the amount of $CO_2$ emission generated if the vehicle is travelling at the speed $s_i(k)$. According to \cite{liu2015distributed}, each cost function is continuous, strict convex and second order differentiable. In particular, $f_i$ has been adopted as an emission function of $s_i$ in an average speed model in the following form:


\begin{equation} \label{avespeed}
	f_i(s_i) = k \left(    \frac{a + bs_i + cs_i^2 + ds_i^3 + es_i^4 + fs_i^5 + gs_i^6}{s_i} \right)
\end{equation} 

\noindent where the parameters $k, a, b, c, d, e, f, g$ are constant values and they are defined differently by different types of ICEVs as per the reference \cite{emfactor}. With this in place, the specific problem to be solved can be formulated in the following:

\begin{equation} \label{eq:opt}
	\begin{gathered}
		\underset{\textbf{s} \in \R^N}{\min} \quad
		\sum\limits_{i\in\underbar{N}} f_{i}\left(s_i \right),\\
		{\text{s.t.}} ~
		s_i = s_j, ~ \forall i \neq j \in \underbar{N}.
	\end{gathered}
\end{equation}


\noindent \textit{Comment:} The ultimate objective of a CSAS is to recommend a virtual consensus speed $\textbf{s}^{*}$ to a group of ICEVs. In the DP-CSAS approach, this is done by solving \eqref{eq:opt} through iterations of $\textbf{s}(k)$. We note that the recommended speed deem to be a virtual speed, and thus it is not our intention to enforce every driver to follow such a speed, which is the key difference between a CSAS and a platooning system. 

\subsection{Related Solutions for CSAS}

SAS is a cooperative system which aims to improve energy-efficient and sustainable mobility for vehicles. Different from the conventional approaches such as the works in~\cite{hounsell2009,tradisauskas2009,gallen2013}, we focus on the consensus problem of SAS in this work. Although consensus problem can be solved in a variety of ways such as using ADMM~\cite{ADMM2}, our focus is to construct a partially distributed solution which allows to calculate an optimal solution in real time. Moreover, different from the works in CSAS designs~\cite{Schakel2013, Ordonez-Hurtado2014}, we address the optimization problem in a privacy-preserving way without revealing individual cost functions of vehicles.

To the best of our knowledge, only a few papers in the literature have considered the privacy issue of the CSAS. Specifically, the work~\cite{griggs2017leader} achieves the consensus over a multi-layer network that no vehicle knows the exact state of other vehicles. Different from that, we focus on protecting the emission cost function of vehicles and our target is to reduce vehicle emissions. The problem studied in the work~\cite{liu2015distributed} is the same as ours. However, as we have described, the proposed DP-CSAS has a few shortcomings. In what follows, we will give a brief review of the DP-CSAS solutions.

\subsection{The State of the Art DP-CSAS Solution}

The key idea of the DP-CSAS solution in \cite{liu2015distributed} is to find the equivalent Lagrange equations for \eqref{eq:opt}, that is:

\begin{equation} \label{kkt}
\frac{\partial \left[ \sum_{i = 1}^{N} f_i(s_i(k)) + \lambda_i (s_i(k) - s_j(k)) \right]}{{\partial s_i(k)}} = 0, ~ \forall i \neq j \in \underbar{N} 
\end{equation}

\noindent where $\lambda_i$ denotes the Lagrange multiplier of the $i$'th constraint. It is not difficult to derive from  \eqref{kkt} that finding the optimal solution of \eqref{eq:opt} is equivalent to solving the following equation:

\begin{equation} \label{eq_opt_new}
\begin{gathered}
\sum_{i=1}^{N} f_i'(s_i) = 0  \\
s_i = s_j, ~ \forall i \neq j \in \underbar{N} \\
\end{gathered}
\end{equation} 

To solve \eqref{eq_opt_new}, an iterative feedback scheme has been applied in the form of:

\begin{equation} \label{iter_eq}
\textbf{s}(k+1) = P(k) \textbf{s}(k) -\mu \sum_{i = 1}^{n} f'_{i}(s_{i}(k)) \textbf{e}
\end{equation}

\noindent where $\{ P(k) \}_{k\in\N} \subset \R^{N \times N}$ is a uniformly strongly ergodic sequence of row stochastic matrices which can be used to model the connectivity among moving vehicles. $\textbf{e} \in \R^{N}$ is a consensus vector with all entries equal to 1.  $u$ is a parameter which determines the convergence and stability of the system. It has been proved in \cite{liu2015distributed} that if $f_i$ are strictly convex, continuous, and each $f'_{i}$ has a strictly positive and bounded growth, i.e. there exist constants  $d_{\min}^{i}$ and $d_{\max}^{i}$ such that for any $a \neq b$

\begin{equation} \label{bdd}
0 < d_{\min}^{i} \leq \frac{f'_{i}(a) - f'_{i}(b)}{a - b} \leq d_{\max}^{i}, ~ \forall i \in \underline{\textrm{N}}. 
\end{equation}

\noindent and when $\mu$ is chosen according to 

\begin{equation} \label{rangemu}
0 < \mu < 2 \left( \sum\limits_{i=1}^n d_{\max}^{i} \right)^{-1}
\end{equation}

\noindent then \eqref{iter_eq} is uniformly globally asymptotically stable at the unique optimal point $\textbf{s}^{*} = s^{*} e$ of the problem (\ref{eq:opt}). 

As a concluding remark of this section, we note that the DP-CSAS proposed in \cite{liu2015distributed} leverages the derivative of each cost function $f_i$, and it requires some strict conditions (Lipschitz condition) on each derivative function $f'_i$ , which might not be practical for cost functions in all ITS applications.

\section{The Proposed Approach}
\label{sec:design}

In this section, we first demonstrate how we can use MPC for privacy-preserving CSAS design with an example. Then, we give the details of our proposed MPC-CSAS.

\subsection{MPC and Its Application to CSAS} \label{mpcapp}

\begin{figure*}[!t]
	\begin{center}
		{\includegraphics[width=0.85\textwidth]{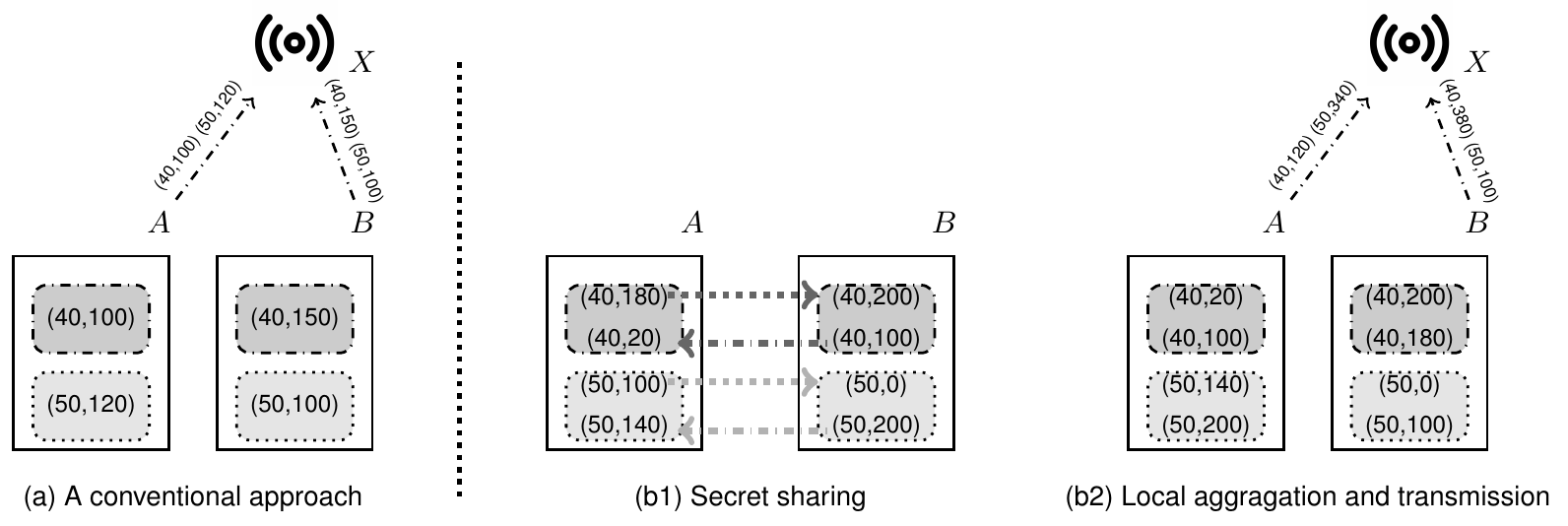}}
		\caption{An example of how to use MPC to consensus-based speed advisory systems.}
		\label{fig:example}
	\end{center}
\end{figure*}

Multi-party Computation (MPC) is a cryptographic functionality that allows 
for secure computation over sensitive data sets~\cite{volgushev2019conclave}. In an MPC protocol, all parties can cooperatively evaluate for some functions, with guarantee that each party can only learn from the output and its own private input. We can construct MPC protocols in different ways, and one of the most commonly used method is the homomorphic secret sharing~\cite{shamir1979share}. In such an approach, the type of the transformation from one algebraic structure into another is kept as the same.  Namely, for any kind of manipulation over the original data, there is a corresponding manipulation of the transformed data~\cite{schoenmakers1999simple}.

To demonstrate how we can apply MPC to the privacy-preserving CSAS design, a simple example with two vehicles,  $A$ and $B$, and a base station $X$, is illustrated in Fig.~\ref{fig:example}. For simplicity, we only include emission values of the two vehicles when speeds are 40 and 50km/h. Here, the objective is to find out which speed can achieve a smaller values. As shown in Fig.~\ref{fig:example}(a), following a conventional approach, we can ask $A$ and $B$ to send the values of their speed-emission mappings to $X$. In this condition, the base station can calculate the sum of the mappings at the speed 40 and 50 respectively, and then recommend the better corresponding speed. For instance, $X$ knows that the sum of the mappings is 250 for the speed 40, and 220 for the speed 50, and clearly 50 will be sent to $A$ and $B$ by broadcast. Obviously, the privacy of $A$ and $B$ is not preserved in this method, since the base station knows their private information. In order to protect private information to the infrastructure (i.e., the base station), we can ask $B$ to send its speed-emission mapping information to $A$ to get an optimal speed. However, $B$ will lose its privacy by revealing its value to $A$ in this setup.

To address this issue, an MPC-based approach is demonstrated in Fig.~\ref{fig:example} (b). The processing mainly includes three steps as follows:

\begin{enumerate}
\item Secret generation: The speed-emission mappings of $A$ and $B$ are split into several shares and some shares should be kept locally. As shown in Fig.~\ref{fig:example} (b1), we use two shares for each mapping, that is the mapping (40,100) is split into (40,180) and (40,20). Note that we doubled the original mapping value here, i.e., from 100 to 200 in this example, but this splitting is not unique. 

\item Secret sharing: For the secret generated for each speed, $A$ and $B$ share part of them to each other.  For example, $A$ shares (40,180) and (50,100) to $B$, and $B$ shares (40,100) and (50,200) to $A$. After that,  the data located on $A$ and $B$ is demonstrated in Fig.~\ref{fig:example} (b2).

\item Local aggregation: $A$ and $B$ aggregate to the local and received speed-emission mappings based on the value of each speed and then send the information to the base station for final speed recommendation. In our example, $A$ sends (40, 120) and (50, 340) to $X$, and $B$ sends (40, 380) and (50, 100).
\end{enumerate}

After the above three steps, $X$ will know that the emission is 500 for the speed 40, and 440 for the speed 50. Therefore, the two vehicles can get the same speed recommendation (i.e., 50km/h) as the conventional approach shown in Fig.~\ref{fig:example}(a). Moreover, since both $A$ and $B$ have kept part of their secret locally, they cannot get any information of the original values from each other. Therefore, the privacy of both $A$ and $B$ is well preserved.

\medskip
\noindent \textit{Comment:} In the above example, for a speed $s_i$ and its emission $f_i(s_i)$, we simply generate two secret values $f_i^{1}(s_i)$ and  $f_i^{2}(s_i)$ on the emission, and guarantee that the functions $f_i^{1}$ and $f_i^{2}$ always meet the condition $f_i(s_i)=0.5 \times (f_i^{1}(s_i) + f_i^{2}(s_i))$ for any $s_i$. This is consistent with the definition of an MPC protocol, since the transformation of algebraic structure is preserved in the assigned problem, i.e.,  $\sum_i f_i(s_i) = 0.5 \times ( \sum_i  f_i^{1}(s_i) + \sum_i  f_i^{2}(s_i) )$.  Therefore, our above operations can achieve the privacy-preserving computation. In contrast, if we generate secrets using another way like $f_i(s_i)= f_i^{1}(s_i) \times f_i^{2}(s_i)$, then the computation will be invalid within the MPC scheme. In general, if $M$ secret values need to be generated for each vehicle, then we can have $g (f_i(s_i)): = a f_i(s_i) +  b =  \sum_{h=1}^{M} f_i^{h}(s_i)$ for any $s_i$, where $a, b$ are constants.

\subsection{The MPC based Privacy-preserving CSAS}

Now we propose the MPC based privacy-preserving CSAS in a formal way by using graph theories. We define an order pair  $\mathcal{G}=(V, E)$ as a directed graph for all vehicles in the network. $V$ denotes the set of  vertices, or vehicles of the graph $\mathcal{G}$. $E$ is a set of edges, or communication links between vehicles, which are order pairs with two distinctive vertices. In our context, this implies that a communication link between any two vehicles in the network is directional. Let $deg^{+}(v)$ denotes the outdegree of a vertex $v \in V$, which models the number of vehicles that can receive a signal from the vehicle represented by the vertex $v$, and let $v^{+}$ be the set of vertices of $v$ which it can reach to. Let $v^{-}$ denote the set of vertices of $v$ which it can receive signal from. We assume that $deg^{+}(v)$ is known to every vertex $v$ in the network. 

With this in place, we claim that for every vehicle, represented by the vertex $v$ in a directed graph $\mathcal{G}$, the MPC based CSAS can preserve privacy for all vehicles in the network if $deg^{+}(v) >= 1, ~v \in V$. Indeed, if the outdegree can hold for every $v \in V$, then it implies that every vehicle in the network can have at least one neighbour to do the secret sharing part shown in Section \ref{mpcapp}, which preserves the privacy of any vehicle by splitting the emission mapping to at least two parts. Now we present the pseudo code of the MPC based privacy-preserving algorithm in the following:

\begin{algorithm}[htbp]
	\caption{MPC-CSAS Privacy-Preserving Algorithm}
	\begin{algorithmic}[1]
		\For{each $i \in \underbar{N}$}
		\State Generate a sequence of $M$ pairs of the original speed-emission mappings in a given range of speeds.
		\State Get the outdegree value for vehicle $i$ corresponding to vertex $v_i$ in the graph $\mathcal{G}$, i.e. $deg^{+}(v_i)$.		
		\For{$j = 1, 2, ..., M$} 
		\State Split the mapping into $deg^{+}(v_i)+1$ shares. 
		\State Reserve one share locally. 
		\State Transmit $deg^{+}(v_i)$ shares to vehicles in the set $v_i^{+}$.
		\EndFor 		
		\State Aggregate the local received mappings from set $v_i^{-}$. 
		\State Send the aggregated local mappings to a base station. 
		\EndFor
		\State Get the best pair index $j^{*}$ from the base station. 
		\State Set the recommended speed to the speed in $j^{*}$th pair.
	\end{algorithmic}
	\label{alg:Algorithm}
\end{algorithm}


\noindent \textit{Remark on Network Connectivity:} Our proposed Algorithm \ref{alg:Algorithm} can be implemented in a batch manner, that is the optimal consensus solution of the problem \eqref{eq:opt} can be found in just one simple algorithm iteration between vehicles and the base station if the outdegree condition for all vehicles can be satisfied in the network. By definition, a strongly connected graph is a graph which for any two vertices $u, v \in V$, there exists a directed path from $u \rightarrow v$ and $v \rightarrow u$, respectively. This show that the proposed MPC-CSAS algorithm is applicable to strongly connected networks. 

On the other hand, vehicular networks are not always static due to various interferences in urban areas. To deal with this challenge, vehicular networks have been modelled using time-varying connectivity graphs in current literature. For instance, in our previous work \cite{liu2015distributed}, a sequence of row-stochastic matrices $\{ P(k) \}_{k\in\N} \subset \R^{N \times N}$ has been applied for this purpose. In particular, if $\{ P(k) \}$ is a uniformly strongly ergodic sequence as the assumption made in \cite{liu2015distributed}, then it implies that every vehicle in the network can communicate to other vehicles over a certain period of time. Roughly speaking, this is equivalent to saying that the resulting connectivity graphs are strongly connected at most time instances \cite{liu2015distributed}, which shows the applicability of our proposed algorithm under switching graph topologies. It is worth noting that in most cases we assume that a group of moving vehicles are relatively static and the distance between each car is relatively small so that the V2V communication is achievable among the group. This assumption is consistent with our previous work \cite{liu2015distributed}. However, our proposed solution can be easily extended to cover the case where a group of vehicles can be dynamically clustered and created according to the quality of network connectivity. In such a case, if a vehicle is driving too fast/slow on a road and it cannot receive the message from a given base station, it is still possible for the vehicle to join a different group of vehicles implementing the MPC-CSAS on another stretch of road.  

Finally, if a vehicle in a group has weakly connectivity to its neighbouring vehicles, it is still possible for the vehicle to conduct the secret sharing and the local aggregation steps with the base station. In this case, a new service needs to be incorporated into the base station which can be seen as a ``dummy'' vehicle for the application purpose. Here, the ``dummy'' vehicle is considered as a software component deployed at a base station, which can assist the weakly connective vehicle for V2V based information exchange. Once this is done, the speed recommendation service, essentially another software component based on the MPC-CSAS algorithm deployed at the base station, can be involved for the final decision making step. From an operational perspective, this setup has flexibility in that it will allow vehicles to conduct a two-step communication with the base station to receive the optimal speed advisory information at different time and place during travelling. We note that although both software components can be deployed at a given base station, the privacy of vehicles' data can still be preserved provided that a security access control based approach is employed at the base station. We shall ignore a further discussion for this aspect in the work as this is beyond the scope of our paper.

\medskip
\noindent \textit{Remark on Network Communication Overheads:} In fact, MPC has significant computational overhead and always involves high communication cost in the presence of big data processing~\cite{volgushev2019conclave}. However, for a CSAS case, the domain of vehicle speeds is small. Moreover, getting an absolute optimal speed will make little sense for CSAS. Therefore, we can ask each vehicle to keep a local list of speed-emission mappings in a specified speed range, such as between 30km/h and 120km/h with increasing every 5km/h. In this case, each vehicle will only need to keep 19 speed-emission pairs locally and shares the 19 pairs with another vehicle. If we use integers to represent the speed and emission, the size of the shared information between vehicles is only 152 bytes. Moreover, the data received and processed by a base station will be less than 3KB for a group of 20 vehicles. Obviously, the amount of data to be processed is extremely small in our approach, and the communication and computing time will be approximately 0 for both vehicles and base stations which are with a general hardware configuration. Namely, the overhead of the MPC is negligible. This is also the reason why we call our approach as a real-time solution for privacy preserved CSAS. Specifically, in our simulation environment, we have observed that with inserted timestamps both the data sharing and data aggregation process can be done in less than 5ms.

\medskip
\noindent \textit{Remark on practical implications of applying the proposed methodology:} We believe that our current proposed method can be easily extended to cope with various setups whilst utilising most up-to-date in-vehicle communication and computing devices as well as roadside traffic infrastructure. Specifically, our proposed approach fits well into the current Mobile Edge Computing (MEC) framework and architecture, where various services can be defined, deployed and implemented at the edge side using for instance REST-based APIs. As we have already commented, our proposed system requires very limited communication and computation overhead (less than 3KB data for 20 vehicles) compared to other existing works in the literature. Given this, we believe that the instantaneity of our system operation in a real world speed advisory scenario is fully operable in current 4G/LTE infrastructures, and will surely be working well in the 5G and beyond 5G network environment.


\section{SUMO Simulation and Results} 
\label{sec:eva}

In this section, we evaluate the performance of Algorithm~\ref{alg:Algorithm} using SUMO \cite{Behrisch2011}. 

\subsection{Simulation Setup}
\begin{figure}[h] 
	\centering
	\includegraphics[width=3.5in]{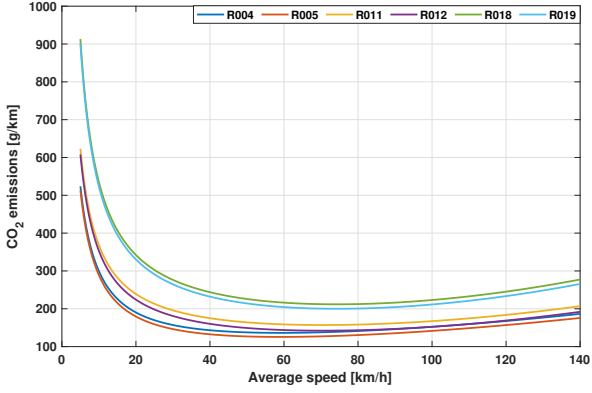}
	\caption{Different CO$_{\mbox{\footnotesize{2}}}$ emission cost functions in Table \ref{tab:Table_emissions}.}
	\label{figco2}
\end{figure}

In order to illustrate the efficacy of our proposed algorithm, we first adopt the average speed model in \eqref{avespeed} for different types of ICEVs with the parameters $a,b,c,d,e,f,g,k$ illustrated in Table \ref{tab:Table_emissions}. The resulting emission functions for six types of ICEVs are shown in Fig. \ref{figco2}.

\begin{table}[h]
	\caption{Emission factors for some CO$_{\mbox{\footnotesize{2}}}$ emission types of ICEVs according to \cite{emfactor}, where $\left\{\text{e},\text{f},\text{g}\right\}=0$ and $\text{k}=1$.}
	\begin{centering}
		\begin{tabular}{|c|c|c|c|c|}
			\hline 
			Type & a & b & c & d \tabularnewline
			\hline 
			\hline 
			R004 & { 2.2606E+3} & { 7.0183E+1} & { 2.9263E-1} & { 3.0199E-3} \tabularnewline
			\hline 
			R005 & { 2.2606E+3} & { 5.9444E+1} & { 2.9263E-1} & { 3.0199E-3} \tabularnewline
			\hline 
			R011 & { 2.5324E+3 } & { 1.1834E+2} & { -4.3167E-1} & { 6.6776E-3} \tabularnewline
			\hline 
			R012 & { 2.5324E+3} & { 1.0340E+2} & { -4.3167E-1} & { 6.6776E-3}
			\tabularnewline
			\hline  
			R018 & { 3.7473E+3} & { 1.6774E+2} & { -8.5270E-1} & { 1.0318E-2}
			\tabularnewline
			\hline
			R019 & { 3.7473E+3} & { 1.5599E+2} & { -8.5270E-1} & { 1.0318E-2}
			\tabularnewline
			\hline
		\end{tabular}
		\par\end{centering}
	\label{tab:Table_emissions}
\end{table}

It is clear to see from Fig. \ref{figco2} that for same emission standard, e.g. Euro 3, increasing engine capacity of the vehicles, i.e. R004, R011 and R018, will also increase the CO$_{\mbox{\footnotesize{2}}}$ amount for a given average speed. 

We now perform the following experiments.

\begin{enumerate}
	\item In the first case, we assume that there are six vehicles, one vehicle in each class as outlined in Table \ref{tab:Table_emissions}. Also, we assume that the topology of the vehicular network is static and strongly connected. In particular, each vehicle is assumed to be connected circle-wise, that is vehicle 1 has one neighbour vehicle 2, vehicle 2 has one neighbour vehicle 3, likewise, up to vehicle 6 which has one neighbour vehicle 1. Finally, we set $M=100$ which linearly spaces the speed in the range of $[5, 140]$km/h, and with the mapping function $g(f_i(s_i)) = f_i(s_i)$. 
	
	\item In the second case, we assume that the system setup is the same to the first case, but with the mapping function $g(f_i(s_i)) = 2 f_i(s_i) + 10$. 
	
	\item In the third case, we assume that there are 20 vehicles for each emission type, and the vehicular network is still static and strongly connected, but with $M=20, 30, \dots, 100$ with the same mapping function in the second case. 

\end{enumerate}

\subsection{Simulation Results}

\begin{figure}[h] 
	\centering
	\includegraphics[width=3.5in]{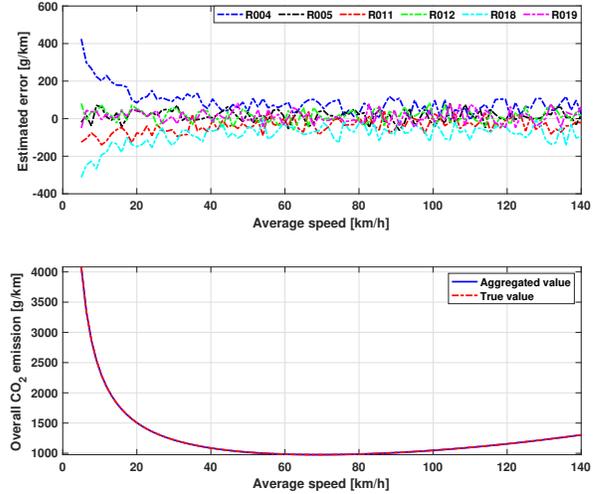}
	\caption{Local estimated error (upper plot) and aggregated curve seen from the base station (lower plot) for the first case.}
	\label{figerror}
\end{figure}

\begin{figure}[h] 
	\centering
	\includegraphics[width=3.5in]{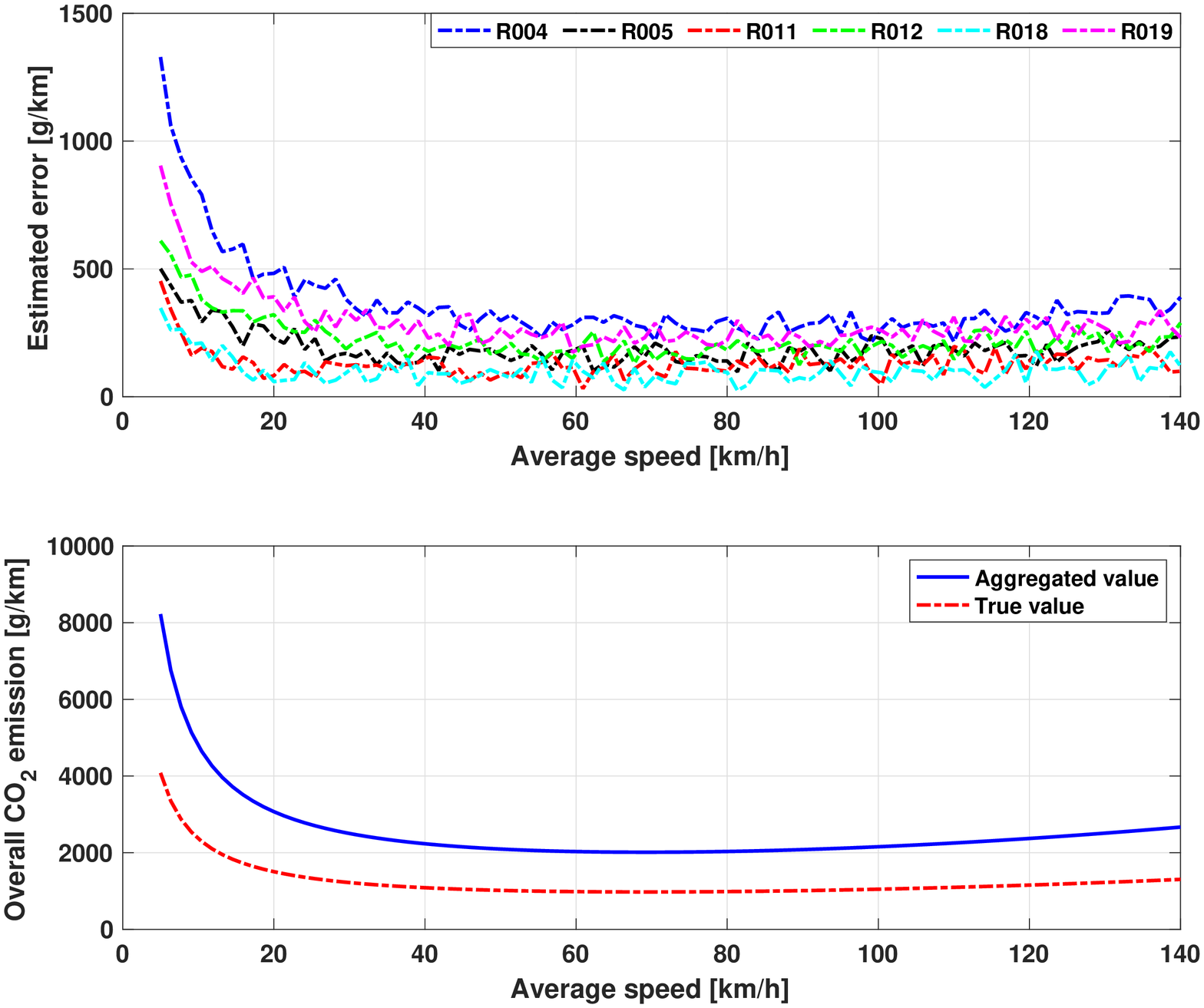}
	\caption{Local estimated error (upper plot) and aggregated curve seen from the base station (lower plot) for the second case.}
	\label{figerror2}
\end{figure}

\begin{figure}[h] 
	\centering
	\includegraphics[width=3.5in]{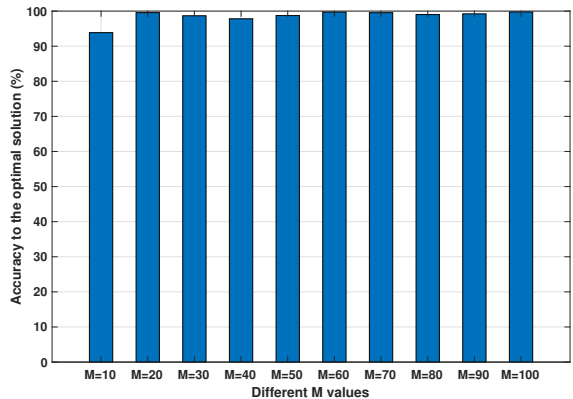}
	\caption{Accuracy of our proposed method compared to the true optimal solutions with respect to $M$.}
	\label{figaccuracy}
\end{figure}

Our simulation results for each individual case study listed above are shown in Figs. \ref{figerror} - \ref{figaccuracy},  respectively. In Fig. \ref{figerror}, we show that the local estimated error from each vehicle has been different from zero, which indicates that no vehicle can correctly identify other vehicles' cost functions by using the secret sharing mechanisms. However, with a basic mapping function $g(x) = x$, the overall cost function seen from a base station is the same with the ground truth, which demonstrates that the privacy of the overall cost function does not preserve well to a central agent in this case. To improve this, we now impose a new linear mapping function $g(x) = 2x + 10$, and it is clear to show from Fig. \ref{figerror2} that privacy for both local and the central agents have been preserved well by using our proposed algorithm while without affecting the optimal solutions of the original problem. Finally, the results in Fig. \ref{figaccuracy} compare the accuracy of optimal solutions by using the proposed algorithm with respect to varying values of $M$ within a set. Clearly, when $M$ is chosen as the lowest value in Fig. \ref{figaccuracy}, i.e. with only 10 data points linearly sampled in the given speed range, the algorithm can perform well by reaching to over 90\% optimality in simple one algorithm iteration. For other cases shown in the figure, we find out that the accuracy is almost identical and nearly 100\% due to the fact that the converged optimal recommended speed is much close to the true optimal value of the original optimization problem. This fact further shows the efficacy to deploy the proposed algorithm in real-time CSAS scenarios, i.e. the value $M$ is not quite sensitive to the accuracy of the optimal solution in our problem. As a final remark, we note that the above simulation results are also applicable to time-varying network having uniformly strongly ergodic property. The stability and convergence of the time-varying graph is guaranteed by our results in \cite{liu2017stability}. The optimality and consistency of the result is guaranteed by the mechanism of the proposed MPC-CSAS algorithm.

\section{Conclusion}  
\label{sec:con}

In this paper, we propose a new design for CSAS based on the MPC protocols, namely MPC-CSAS. We have shown that by using the ideas of secret generation and sharing, users' privacy can be designed and preserved both locally and globally for a CSAS application, that is no other vehicles or a base station in the network can infer the original cost function associated with each vehicle, which may contain a user's critical/sensitive information. We have discussed the applicability and feasibility to deploy the MPC-CSAS with both static and strongly connected network as well as time-varying network which satisfies uniformly strongly ergodic property linked to graph theories. We have validated the efficacy of the proposed algorithm in SUMO simulations. One of the assumptions we made in this paper is that a participated vehicle can trust the information received from other vehicles in the network, which may potentially cause security concerns to the participated vehicle. In light of this, we will explore trustiness detection strategies such as how to use blockchain to record trustiness of the involved vehicles as part of our future work. We will also explore how the proposed MPC-CSAS can be adapted to more complex networks for novel scenarios in ITS.

\bibliographystyle{ieeetran}
\bibliography{IEEEabrv,mpc}

\end{document}